\documentclass[11pt,leqno]{amsart}
\usepackage{amsmath}
\usepackage{amssymb}
\begin{document}

\newcommand{\defi}{\stackrel{\Delta}{=}}
\newcommand{\B}{\mathcal{B}}
\newcommand{\U}{\mathcal{U}}
\newcommand{\G}{\mathcal{G}}
\newcommand{\cZ}{\mathcal{Z}}
\newcommand\one{\hbox{1\kern-2.4pt l }}
\newcommand{\Item}{\refstepcounter{Ictr}\item[(\theIctr)]}
\newcommand{\QQ}{\hphantom{MMMMMMM}}

\newtheorem{Theorem}{Theorem}[section]
\newtheorem{Lemma}[Theorem]{Lemma}
\newtheorem{Corollary}[Theorem]{Corollary}
\newtheorem{Remark}[Theorem]{Remark}
\newtheorem{Proposition}[Theorem]{Proposition}
\newtheorem{Definition}[Theorem]{Definition}
\newtheorem{Construction}[Theorem]{Construction}
\newtheorem{Example}[Theorem]{Example}

\newcounter{claim_nb}[Theorem]
\setcounter{claim_nb}{0}
\newtheorem{claim}[claim_nb]{Claim}
% New list environment
\newenvironment{cproof}
{\begin{proof}
 [Proof.]
 \vspace{-3.2\parsep}}
{\renewcommand{\qed}{\hfill $\Diamond$} \end{proof}}

\newcounter{Ictr}

\renewcommand{\theequation}{%\thesection.
\arabic{equation}}

\def\A{\mathcal{A}}

\def\C{\mathcal{C}}

\def\V{\mathcal{V}}

\def\I{\mathcal{I}}

\def\Y{\mathcal{Y}}

\def\X{\mathcal{X}}

\def\J{\mathcal{J}}

\def\Q{\mathcal{Q}}

\def\W{\mathcal{W}}

\def\S{\mathcal{S}}

\def\T{\mathcal{T}}

\def\L{\mathcal{L}}

\def\M{\mathcal{M}}

\def\N{\mathcal{N}}
\def\P{\mathcal{P}}
\def\R{\mathbb{R}}
\def\H{\mathbb{H}}

\def\Diag{\textup{Diag}}
\def\trace{\textup{Tr}}

%%%%%%%%%%%%%%%%%%%%%%%%%%%%%%%%%%%%%%%%%%%%%%%%%%%%%%%%%%%%%%%%%%%%%%%%

\title{Sufficient Conditions for Low-rank Matrix Recovery,\\ Translated from\\  Sparse Signal Recovery}


\author{Lingchen Kong, Levent Tun\c{c}el, Naihua Xiu}
\thanks{Lingchen Kong: Department of Applied Mathematics, Beijing Jiaotong
University, Beijing 100044, P. R. China (e-mail:
konglchen@126.com)\\
Levent Tun\c{c}el: Department of Combinatorics and Optimization, Faculty
of Mathematics, University of Waterloo, Waterloo, Ontario N2L 3G1,
Canada (e-mail: ltuncel@math.uwaterloo.ca)\\
Naihua Xiu: Department of Applied Mathematics,
Beijing Jiaotong University, Beijing 100044, P. R. China
(e-mail: nhxiu@bjtu.edu.cn)}

\date{June 15, 2011}

\begin{abstract} The low-rank matrix recovery (LMR) is a rank minimization problem
subject to linear equality
constraints, and it arises in many fields such as signal and image
processing, statistics, computer vision,
system identification and control. This class of
optimization problems is $\N\P$-hard and a popular approach
replaces the rank function with the nuclear norm of the matrix
variable. In this paper, we extend
the concept of $s$-goodness for a sensing matrix in
sparse signal recovery (proposed
by Juditsky and Nemirovski [Math Program, 2011])
to linear transformations in LMR.  Then, we give
characterizations of $s$-goodness in the context of LMR.
Using the two characteristic $s$-goodness constants,
${\gamma}_s$ and $\hat{\gamma}_s$,
of a linear transformation, not only do we
derive necessary and sufficient conditions
for a linear transformation to be $s$-good, but also
provide sufficient conditions for exact and stable
$s$-rank matrix recovery via the nuclear norm minimization
under mild assumptions.
Moreover, we give computable upper bounds for one
of the $s$-goodness
characteristics
which leads to verifiable sufficient
conditions for exact low-rank matrix recovery.
\end{abstract}


\maketitle

\section{Introduction}
The \emph{low-rank matrix recovery (LMR for short)} is a \emph{rank minimization problem (RMP)}
with linear constraints, or the \emph{affine matrix rank minimization problem} which is defined as follows:
\begin{eqnarray}\label{1f1} \textup{minimize} ~{\rm rank}(X),\ \ \
\textup{subject to}~~\A X=b,\end{eqnarray} where $X\in\R^{m\times n}$ is the
matrix variable, and $\A: \R^{m \times n}\rightarrow \R^p$ is a
linear transformation and $b\in\R^p$.
 Although specific instances can often be solved with specialized algorithms,
the LMR is $\N\P$-hard.  A popular approach for solving LMR
in the systems and control community is to minimize the trace of a positive semidefinite matrix variable
instead of the rank (see, e.g., \cite{BD98,MP97}). A generalization
of this approach to non-symmetric matrices introduced by Fazel,
Hindi and Boyd \cite{FHB01} is the famous convex relaxation of LMR
(\ref{1f1}), which is called \emph{nuclear norm minimization
(NNM)}:
\begin{eqnarray}\label{1f1-}\min ~\|X\|_\ast \ \ \
{\rm s.t.}~~\A X=b,\end{eqnarray} where $\|X\|_\ast$ is the \emph{nuclear
norm} of $X$, i.e., the sum of its singular values. When $m=n$ and
the matrix $X:={\Diag}(x), x\in \R^n$, is diagonal, the
LMR (\ref{1f1}) reduces to \emph{sparse signal recovery (SSR)}, which
is the so-called \emph{cardinality minimization problem (CMP)}:
\begin{eqnarray}\label{1f2}\min ~\|x\|_0 \ \ \
{\rm s.t.}~~\Phi x=b,\end{eqnarray} where $\|x\|_0$ denotes the
number of nonzero entries in the vector $x$, $\Phi\in \R^{m \times
n}$ is a sensing matrix. A well-known heuristic for SSR is  the
\emph{$\ell_1$-norm minimization} relaxation (basis pursuit problem):
\begin{eqnarray}\label{1f2-}\min ~\|x\|_1 \ \ \
{\rm s.t.}~~\Phi x=b,\nonumber\end{eqnarray} where $\|x\|_1$ is the
\emph{$\ell_1$-norm} of $x$, i.e., the sum of absolute values of its entries.

The LMR problems have many applications and appeared in the literature of a
diverse set of fields including signal and image processing,
statistics, computer vision, system identification and control. For more details,
see the recent survey paper \cite{RFP08}. LMR and NNM have been the
focus of some recent research in optimization community, see, e.g.,
\cite{AV09,CCS10,CR09,DST10,LCWM09,LST09,LV09,MGC10,PGWXM10,RFP08,RXH08,TY11}. Although there are many
papers dealing with algorithms for NNM such as interior-point
methods, fixed point and Bregman iterative methods and proximal
point methods, there are very few papers dealing with the conditions that
guarantee the success of the low-rank matrix recovery via NNM. For
instance, following the program laid
out in the work of Cand\`{e}s and Tao in compressed sensing (CS,
see, e.g., \cite{CRT06,CT05,D06}), Recht, Fazel and Parrilo \cite{RFP08}
provided a certain \emph{restricted isometry property} (RIP) condition
on the linear
transformation which guarantees the minimum nuclear norm
solution is the minimum rank solution.
Recht, Xu and Hassibi \cite{RXH08, RXH2011} gave another
condition which characterizes a particular property of the
null-space of the linear transformation.

In the setting of CS, there are other characterizations of the
sensing matrix, under which $\ell_1$-norm minimization can be
guaranteed to yield an optimal solution to SSR, in addition to RIP and
null-space properties, see, e.g., \cite{DG08,JN08,JKN08,JKN10}. In particular, Juditsky and Nemirovski
\cite{JN08} established necessary and sufficient conditions for a
sensing matrix to be ``\emph{$s$-good}" to allow for exact $\ell_1$-recovery
of sparse signals with $s$ nonzero entries when no measurement noise
is present. They also demonstrated that these characteristics,
although difficult to evaluate, lead to verifiable
sufficient conditions for exact SSR and to efficiently computable
upper bounds on those $s$ for which a given sensing matrix is $s$-good. Furthermore, they established
instructive links between $s$-goodness and RIP in the CS context.
One may wonder whether we can generalize the
$s$-goodness concept to LMR and still maintain
many of the nice properties as done in \cite{JN08}.
Here, we deal with this issue. Our approach is based on the  singular value
decomposition (SVD) of a matrix and the partition technique generalized from CS.
In the next section, following Juditsky and Nemirovski's terminology, we propose
definitions of $s$-goodness and $G$-numbers of a linear
transformation in LMR. We provide some basic properties of $G$-numbers. In Section 3, we characterize $s$-goodness of
a linear transformation in LMR via $G$-numbers.
We establish the exact and stable LMR results in Section 4.
In Section 5, we show that these characteristics lead to verifiable sufficient conditions for
exact $s$-rank matrix recovery and to computable upper bounds on
those $s$, %via a semi-infinite programming problem,
for which a given linear
transformation is $s$-good. In Section 6, we consider the connection between
$s$-goodness  and RIP for a linear transformation in LMR. As a byproduct, we obtain the new bound on restricted isometry constant
$\delta_{2s}<\sqrt{2}-1$. As we were in the final stages of the preparation of this paper,
Oymak, Mohan, Fazel and Hassibi \cite{OMFH2011}
proposed a general technique for translating results from SSR
to LMR, where they give the current best bound on
the restricted isometry constant $\delta_{2s}<0.472$.
These results were independently obtained.
A difference between the results is that
we follow Juditsky and Nemirovski's geometric, optimization
based approach.

Let $W\in\R^{m\times n}, r:=\min\{m, n\}$ and let
$W=U\Diag(\sigma(W))V^T$ be the \emph{SVD} of $W$,
where $U\in \R^{m\times r}, V\in \R^{n\times r}$, and
$\Diag(\sigma(W))$ is the diagonal matrix of
$\sigma(W)=(\sigma_1(W),\ldots,\sigma_r(W))^T$ which is the
vector of the singular values of $W$.
Also let $\Xi(W)$ denote the set of pairs of matrices
$(U,V)$ in the SVD of $W$, i.e.,
$$\Xi(W):=\{(U,V):  U\in \R^{m\times r},
V\in \R^{n\times r}, W=U\Diag(\sigma(W))V^T\}.$$
For $s\in\{0,1,2,\ldots,r\}$, we say $W\in\R^{m\times n}$
is a \emph{$s$-rank matrix} to mean that the rank of $W$ is
no more than $s$. For a $s$-rank matrix $W$, it is convenient to take
$W=U_{m\times s}W_sV^T_{n\times s}$ as its SVD where $U_{m\times s}\in\R^{m\times s}, V_{n\times
s}\in\R^{n\times s}$ are orthogonal matrices and
$W_s=\Diag((\sigma_1(W),\ldots,\sigma_s(W))^T)$. For a vector
$y\in\R^p$, let $\|\cdot\|_d$ be the \emph{dual norm} of $\|\cdot\|$
specified by $\|y\|_d:=\max_v\{\langle v,y\rangle: \|v\|\leq 1\}.$
In particular, $\|\cdot\|_\infty$ is the dual norm of $\|\cdot\|_1$
for a vector. Let $\|X\|$ denote the \emph{spectral or the operator
norm} of a matrix $X\in\R^{m\times n}$, i.e., the largest singular
value of $X$. In fact, $\|X\|$ is the dual norm of $\|X\|_\ast$. Let $\|X\|_F:=\sqrt{\langle
X,X\rangle}=\sqrt{{\trace}(X^TX)}$ be the \emph{Frobenius norm} of $X$,
which is equal to the $\ell_2$-norm of the vector of its singular values. We
denote by $X^T$ the \emph{transpose} of $X$. For a linear transformation
$\A:\R^{m\times n}\rightarrow \R^p$,
we denote by $\A^\ast:\R^p \rightarrow \R^{m \times n}$
the \emph{adjoint} of $\A$.

\section{Preliminaries}
%\section{$S$-goodness and $G$-numbers}
\subsection{Definitions}
We first go over some concepts related to $s$-goodness of the linear
transformation in LMR (RMP).  These are extensions of
those given for SSR (CMP) in \cite{JN08}.

\begin{Definition} Let $\A:\R^{m\times n}\rightarrow \R^p$ be a
linear transformation
and  $s\in\{0,1,2,\ldots,r\}$. We say that
\emph{$\A$ is $s$-good},
if for every $s$-rank matrix $W\in\R^{m\times n}$, $W$ is the unique
optimal solution to the optimization problem
\begin{eqnarray}\label{2f2}
{\min}_{X\in\R^{m\times n}}\{\|X\|_\ast: \A X=\A W\}.\end{eqnarray}
\end{Definition}

We denote by $s_\ast(\A)$ the largest integer
$s$ for which $\A$ is $s$-good.
Clearly, $s_\ast(\A) \in \{0,1, \ldots, r\}.$
To characterize $s$-goodness we introduce two useful
\emph{$s$-goodness constants}: ${\gamma}_s$ and
$\hat{\gamma}_s$, we call ${\gamma}_s$ and
$\hat{\gamma}_s$ $G$-numbers.

\begin{Definition} Let $\A:\R^{m\times n}\rightarrow \R^p$ be a linear
transformation, $\beta\in [0,+\infty]$ and
$s\in\{0,1,2,\ldots,r\}$.  Then,

(i) $G$-number $\gamma_s(\A,\beta)$ is the infimum of $\gamma\geq 0$ such that
for every matrix $X\in \R^{m\times n}$ with singular value
decomposition $X=U_{m\times s}V^T_{n\times s}$ (i.e., $s$ nonzero singular
values, all equal to 1), there exists a vector $y\in\R^p$ such that
\begin{eqnarray}\label{2f3} \|y\|_d\leq \beta {\rm~and~} \A^\ast
y=U \Diag(\sigma(\A^\ast y))V^T,
\end{eqnarray}${where }~ U=[U_{m\times
s}~U_{m\times (r-s)}], V=[V_{n\times s}~V_{n\times (r-s)}]~{
are~orthogonal~matrices,~and}$ $$\sigma_i(\A^\ast y) \left
\{\begin{array}{cc} =1,  &
\textrm{if}~~\sigma_i(X)=1,\cr\noalign{\vskip2truemm}
\in[0,\gamma],  & \textrm{if}~~\sigma_i(X)=0,\\
\end{array}\right.~~i\in\{1,2,\ldots,r\}.$$
If  there does not exist such $y$ for some $X$ as above, we set
$\gamma_s(\A,\beta)=+\infty$.

(ii) ${G}$-number $\hat{\gamma}_s(\A,\beta)$ is the infimum of $\gamma\geq 0$
such that for every matrix $X\in \R^{m\times n}$ with $s$ nonzero
singular values, all equal to $1$, there exists a vector $y\in\R^p$ such
that
\begin{eqnarray}\label{2f4} \|y\|_d\leq \beta {\rm~and~}
\|\A^\ast y-X\|\leq\gamma.
\end{eqnarray}
If  there does not exist such $y$ for some $X$ as above, we set
$\gamma_s(\A,\beta)=+\infty$ and to be compatible with the
special case given by \cite{JN08}, we write $\gamma_s(\A)$,
$\hat{\gamma}_s(\A)$ instead of $\gamma_s(\A, +\infty)$,
$\hat{\gamma}_s(\A,+\infty)$, respectively.
\end{Definition}

From the above definition, we easily
see that the set of values
that $\gamma$ takes is closed.
Thus, when $\gamma_s(\A,\beta)<+\infty$, for
every matrix $X\in \R^{m\times n}$ with $s$ nonzero singular values,
all equal to 1, there exists a vector $y\in\R^p$ such that
\begin{eqnarray}\label{2f5} \|y\|_d\leq \beta {\rm~and~}
\sigma_i(\A^\ast y) \left \{\begin{array}{cc} =1,  &
\textrm{if}~~\sigma_i(X)=1,\cr\noalign{\vskip2truemm}
\in[0,\gamma_s(\A,\beta)],  & \textrm{if}~~\sigma_i(X)=0, \\
\end{array}\right.~~i\in\{1,2,\ldots,r\}.
\end{eqnarray}
Similarly, for every matrix $X\in \R^{m\times n}$ with $s$ nonzero
singular values, all equal to $1$, there exists a vector $\hat{y}\in\R^p$
such that
\begin{eqnarray}\label{2f6} \|\hat{y}\|_d\leq \beta {\rm~and~}
\|\A^\ast \hat{y}-X\|\leq\hat{\gamma}_s(\A,\beta).
\end{eqnarray}
Observing that the set $\{\A^\ast y: \|y\|_d\leq \beta\}$ is convex,
we obtain that if $\gamma_s(\A,\beta)<+\infty$, then
for every matrix $X$ with at most $s$
nonzero singular values and $\|X\|\leq 1$
there exist
vectors $y$
satisfying (\ref{2f5}) and there exist vectors
$\hat{y}$ satisfying (\ref{2f6}).
Moreover, for a given pair $\A$, $s$,
$\gamma_s(\A,\beta)=\gamma_s(\A)$ and
$\hat{\gamma}_s(\A,\beta)=\hat{\gamma}_s(\A)$,
for all $\beta$ large enough. However, we would not want
$\beta$ to be very large in some situations,
see Section 4. Thus, we need to work out an
answer to the question
``what is large enough" in our context. Below, we
give a simple result
in this direction as it was done in the vector case,
see Proposition 2 in
\cite{JN08} for details.

\begin{Proposition}\label{fp1}
Let  $\A:\R^{m\times n}\rightarrow \R^p$ be a linear
transformation and $\beta\in [0,+\infty]$.
Assume that for some $\rho>0$,
the image of the unit $\|\cdot\|_\ast$-ball in
$\R^{m\times n}$ under the mapping $X\mapsto\A X$ contains
the ball $B=\{x\in\R^p: \|x\|_1\leq\rho\}$.
Then for every $s\in\{1,2,\ldots,r\}$,
$$\beta\geq \frac{1}{\rho}
\textup{ and } \gamma_s(\A)<1
~~\Rightarrow~~{\gamma}_s (\A, \beta)= {\gamma}_s (\A).$$
\end{Proposition}
\textbf{Proof.} Fix $s\in\{1,2,\ldots,r\}$. Let $\gamma:=\gamma_s(\A)<1$. Then
for every matrix $W\in \R^{m\times n}$ with its SVD $W=U_{m\times s}V^T_{n\times s}$, there exists a vector $y\in\R^p$ such that
\begin{eqnarray}\label{2p3} \|y\|_d\leq \beta {\rm~and~} \A^\ast
y=U \Diag(\sigma(\A^\ast y))V^T,~\nonumber
\end{eqnarray}${where }~ U=[U_{m\times
s}~U_{m\times (r-s)}], V=[V_{n\times s}~V_{n\times (r-s)}]~{
are~orthogonal~matrices,~and}$ $$\sigma_i(\A^\ast y) \left
\{\begin{array}{cc} =1,  &
\textrm{if}~~\sigma_i(W)=1,\cr\noalign{\vskip2truemm}
\in[0,\gamma],  & \textrm{if}~~\sigma_i(W)=0,\\
\end{array}\right.~~i\in\{1,2,\ldots,r\}.$$Clearly, $\|\A^\ast y\|\leq1$. That is,
$$1\geq \|\A^\ast y\|=\max_{X\in\R^{m\times n}}\{\langle X,\A^\ast y\rangle: \|X\|_\ast\leq 1\}
=\max_{X\in\R^{m\times n}}\{\langle u, y\rangle: u=\A X,\|X\|_\ast\leq 1\}.$$
From the inclusion assumption, we obtain that
$$\max_{X\in\R^{m\times n}}\{\langle u, y\rangle: u=\A X, \|X\|_\ast\leq 1\}
\geq\max_{u\in\R^p}\{\langle u, y\rangle: \|u\|_1\leq \rho\}=\rho\|y\|_\infty=\rho\|y\|_d.$$
Combining the above two strings of relations,
we derive the desired conclusion.
{\qed}

\subsection{Convexity and monotonicity of $G$-numbers}
In order to characterize the $s$-goodness of a linear transformation $\A$,
we study convexity and monotonicity
properties of $G$-numbers.
We begin with the result that $G$-numbers
$\gamma_s(\A,\beta)$ and $\hat{\gamma}_s(\A,\beta)$ are
convex nonincreasing functions of
$\beta$.

\begin{Proposition}\label{ft01}
For every linear transformation $\A$ and every
$s \in \{0,1, \ldots, r\}$, $G$-numbers
$\gamma_s(\A,\beta)$ and $\hat{\gamma}_s(\A,\beta)$ are
convex nonincreasing functions of
$\beta \in [0,+\infty]$.
\end{Proposition}
\textbf{Proof.} We only need to demonstrate that the quantity
$\gamma_s(\A,\beta)$ is a convex nonincreasing function of
$\beta \in [0,+\infty]$. It is evident from the definition
that
$\gamma_s(\A,\beta)$ is nonincreasing for given $\A,s$. It remains
to show that $\gamma_s(\A,\beta)$ is a convex function of $\beta$.
In other words, for every pair $\beta_1,\beta_2\in[0,+\infty]$, we need to
verify that
$$\gamma_s(\A,\alpha\beta_1+(1-\alpha)\beta_2)
\leq\alpha\gamma_s(\A,\beta_1)+(1-\alpha)\gamma_s(\A,\beta_2),
~~\forall \alpha\in[0,1].$$
The above inequality holds immediately if one of $\beta_1, \beta_2$ is
$+\infty$.  Thus, we may assume
$\beta_1,\beta_2\in[0,+\infty)$. In fact, from the argument
around (\ref{2f5}) and the definition of $\gamma_s(\A,\cdot)$, we
know that for every matrix $X=U\Diag(\sigma(X))V^T$
with $s$ nonzero singular values, all equal to
$1$, there exist vectors $y_1, y_2\in\R^p$ such that
for $k \in \{1,2\}$,
\begin{eqnarray} \label{2f0}\|y_k\|_d\leq \beta_k {\rm~and~}
\sigma_i(\A^\ast y_k) \left \{\begin{array}{cc} =1,  &
\textrm{if}~~\sigma_i(X)=1,\cr\noalign{\vskip2truemm}
\in[0,\gamma_s(\A,\beta_k)],  & \textrm{if}~~\sigma_i(X)=0, \\
\end{array}\right.~~i\in\{1,2,\ldots,r\}.
\end{eqnarray}
It is immediate from
(\ref{2f0}) that $\|\alpha y_1+(1-\alpha)y_2\|_d
\leq\alpha\beta_1+(1-\alpha)\beta_2$. Moreover, from the above
information on the singular values
of $\A^\ast y_1, \A^\ast y_2$, we may set
$\A^\ast y_k=X+Y_k$, $k \in \{1,2\}$
such that
$$X^T Y_k=0, XY_k^T=0,~{\rm rank}(Y_k)\leq r-s,
{\rm~and~} \|Y_k\|\leq\gamma_s(\A,\beta_k).$$
This implies for every $\alpha \in [0,1]$,
\[X^T \left[\alpha Y_1+(1-\alpha)Y_2\right]=0,
X\left[\alpha Y_1+(1-\alpha) Y_2\right]^T=0,
\]
and hence ${\rm rank}\left[\alpha Y_1+(1-\alpha)Y_2\right] \leq r-s$,
$X$ and $\left[\alpha Y_1+(1-\alpha)Y_2\right]$ share the same orthogonal
row and column spaces. Thus, noting that $\A^\ast \left[\alpha
y_1+(1-\alpha)y_2\right]=X+\alpha Y_1+(1-\alpha)Y_2$, we obtain that $\|\alpha y_1+(1-\alpha)y_2\|_d
\leq\alpha\beta_1+(1-\alpha)\beta_2 $ and~
\begin{eqnarray}
\sigma_i(\A^\ast
(\alpha y_1+(1-\alpha)y_2))= \left \{\begin{array}{cc} 1,  &
\textrm{if}~~\sigma_i(X)=1,\cr\noalign{\vskip2truemm}
\sigma_i(\alpha Y_1+(1-\alpha)Y_2),  & \textrm{if}~~\sigma_i(X)=0, \\
\end{array}\right.
\nonumber\end{eqnarray}
for every $\alpha \in [0,1]$.  Combining this
with the fact
$$\|\alpha Y_1+(1-\alpha)Y_2\|\leq \alpha \|Y_1\|+(1-\alpha)\|Y_2\|\leq
\alpha\gamma_s(\A,\beta_1)+(1-\alpha)\gamma_s(\A,\beta_2),$$
we obtain the desired conclusion. {\qed}

The following observation that $G$-numbers
$\gamma_s(\A,\beta), \hat{\gamma}_s(\A,\beta)$ are nondecreasing in
$s$ is immediate.

\begin{Proposition}\label{ft02} For every
$s'\leq s$, we have $\gamma_{s'}(\A,\beta)\leq\gamma_s(\A,\beta),
~~ \hat{\gamma}_{s'}(\A,\beta)\leq\hat{\gamma}_s(\A,\beta)$.
\end{Proposition}

%\subsection{Further properties of $G$-number}
%Equivalent representation of $\hat{G}$-number}

We further investigate the relationship between the
$G$-numbers $\gamma_s(\A,\beta)$ and $\hat{\gamma}_s(\A,\beta)$.
The following result generalizes the second part
of Theorem 1 of \cite{JN08} (and its proof).

\begin{Proposition}\label{ft3}
Let  $\A:\R^{m\times n}\rightarrow \R^p$ be a linear
transformation, $\beta\in [0,+\infty]$ and $s\in\{0,1,2,\ldots,r\}$.
Then we have
\begin{eqnarray}\label{2f9} \gamma:=\gamma_s(\A,\beta)<1
~&\Rightarrow &\hat{\gamma}_s
\left(\A,\frac{1}{1+\gamma}\beta\right)=\frac{\gamma}{1+\gamma}<\frac{1}{2};\\
\label{2f10}\hat{\gamma}:=\hat{\gamma}_s(\A,\beta)<\frac{1}{2}
~&\Rightarrow
&\gamma_s
\left(\A,\frac{1}{1-\hat{\gamma}} \beta\right)
=\frac{\hat{\gamma}}{1-\hat{\gamma}}<1.\end{eqnarray}
\end{Proposition}
\textbf{Proof.} Let $\gamma:=\gamma_s(\A,\beta)<1$. Then, for every
matrix $Z\in \R^{m\times n}$ with $s$ nonzero singular values, all equal
to 1, there exists $y\in\R^p$, $\|y\|_\ast\leq \beta$, such that
$\A^\ast y=Z+W$, where $\|W\|\leq \gamma$ and $W$ and $Z$ share the
same orthogonal row and column spaces.
For a given pair $Z,y$ as above, take
$\tilde{y}:=\frac{1}{1+\gamma}y$. Then we have
$\|\tilde{y}\|_\ast\leq \frac{1}{1+\gamma}\beta$ and $$\|\A^\ast
\tilde{y}-Z\|\leq \max\left\{1-\frac{1}{1+\gamma},
\frac{\gamma}{1+\gamma}\right\}=\frac{\gamma}{1+\gamma},$$
where the first term under the maximum comes from the
fact that $\A^\ast y$ and $Z$ agree on the subspace corresponding
to the nonzero singular values of $Z$. Therefore, we
obtain
\begin{eqnarray}\label{2f11}\hat{\gamma}_s
\left(\A,\frac{1}{1+\gamma}\beta\right)
\leq\frac{\gamma}{1+\gamma}<\frac{1}{2}.\end{eqnarray}
Now, we assume that $\hat{\gamma}:=\hat{\gamma}_s(\A,\beta)<1/2$. Fix
orthogonal matrices  $U\in\R^{m\times r}, V\in\R^{n\times r}$. For
an $s$-element subset $J$ of the index set $\{1,2,\ldots,r\}$, we
define a set $S_J$
with respect to orthogonal matrices $U, V$ as
$$S_J:=\left\{x\in \R^r: \exists y\in\R^p,
\|y\|_d\leq \beta,~\A^\ast y=U{\rm \Diag}(\sigma(\A^\ast y))V^T~{\rm
where}~\sigma_i(\A^\ast y) \left \{\begin{array}{cc} =|x_i|, &
\textrm{if}~~i\in J,\cr\noalign{\vskip2truemm}
\leq \hat{\gamma},  & \textrm{if}~~i\in \bar{J}. \\
\end{array}\right.\right\}.$$
In the above, $\bar{J}$
denotes the complement of $J$. It is immediately seen that $S_J$ is a
closed convex set in $\R^r$. As in the proof of Theorem 1
in \cite{JN08}, we have
\begin{claim}
$S_J$ contains the
$\|\cdot\|_\infty$-ball of radius $(1-\hat{\gamma})$ centered at the
origin in
$\R^r$.
\end{claim}

\begin{cproof}
Note that $S_J$ is closed and convex.  Moreover,
$S_J$ is the direct
sum of its projections onto the pair of subspaces
\[
L_J:=\{x\in \R^r: x_i=0,i\in \bar{J}\} \mbox{ and its orthogonal
complement } L_J^\bot=\{x\in\R^r: x_i=0,i\in J\}.
\]
Let $Q$ denote the projection of $S_J$ onto $L_J$.
Then,
$Q$ is closed and convex (because of the direct sum property above
and the fact that $S_J$ is closed and convex). Note that $L_J$ can be naturally
identified with $\R^s$, and our claim is
the image $\bar{Q}\subset\R^s$ of $Q$ under this identification
contains the $\|\cdot\|_\infty$-ball $B_s$ of
radius $(1-\hat{\gamma})$ centered at the origin
in $\R^s$.
For a contradiction, suppose $B_s$ is not
contained in $\bar{Q}$. Then there exists
$v\in
B_s\setminus\bar{Q}$. Since $\bar{Q}$
is closed and convex,
by a separating
hyperplane theorem, there exists a vector $u\in\R^s$,
$\|u\|_1 = 1$ such that $$u^T v
>  u^T v' \mbox{ for every } v' \in \bar{Q}.$$
Let $z\in\R^r$ be defined by
\[
z_i := \left\{\begin{array}{rl}
1, & i \in J,\\
0, & \mbox{otherwise.}
\end{array}
\right.
\]
By definition of $\hat{\gamma}=\hat{\gamma}_s(\A,\beta)$, for $s$-rank
matrix $U{\rm \Diag}(z)V^T$, there exists $y\in\R^p$ such that
$\|y\|_d\leq \beta$ and $$\A^\ast y=U{\rm \Diag}(z)V^T+W,$$ where $W$
and $U \Diag(z)V^T$ have the same row and column
spaces, $\|\A^\ast y- \Diag(z)\|\leq \hat{\gamma}$ and
$\|\sigma(\A^\ast y)-z\|_\infty\leq \hat{\gamma}$. Together with
the definitions of $S_J$ and $\bar{Q}$, this means that $\bar{Q}$
contains a vector $\bar{v}$ with $|\bar{v}_i-\textup{sign}(u_i)|\leq \hat{\gamma}$,
$\forall i \in \{1,2, \ldots,s\}$.
Therefore,
\[
u^T\bar{v}\geq
\sum_{i=1}^s |u_i| (1-\hat{\gamma})
= (1-\hat{\gamma})\|u\|_1=1-\hat{\gamma}.
\]
By $v\in
B_s$ and the definition of $u$, we obtain
\[
1-\hat{\gamma}\geq
\|v\|_\infty = \|u\|_1\|v\|_{\infty}
\geq u^Tv>u^T\bar{v}\geq 1-\hat{\gamma},
\]
where the strict inequality follows from the facts
that $\bar{v} \in \bar{Q}$ and $u$ separates $v$
from $\bar{Q}$.
The above string of inequalities is a
contradiction, and hence the desired claim holds.
\end{cproof}

Using the above claim, we
conclude that for every $J \subseteq \{1,2, \ldots, r\}$ with
cardinality $s$,
there exists an $x\in S_J$ such that $x_i =
(1-\hat{\gamma}), \forall i\in J$.
From the definition of $S_J$, we
obtain that there exists $y\in \R^p$ with $\|y\|_d\leq
(1-\hat{\gamma})^{-1}\beta$ such that
$$\A^\ast y=U{\Diag}(\sigma(\A^\ast y))V^T,$$
where $\sigma_i(\A^\ast
y)=(1-\hat{\gamma})^{-1}x_i=1$ if~$i\in J$, and $\sigma_i(\A^\ast
y)_i\leq (1-\hat{\gamma})^{-1}\hat{\gamma}$ if~$i\in \bar{J}.$ Thus,
we obtain that
\begin{eqnarray}\label{2f12}\hat{\gamma}_s:=\hat{\gamma}_s
(\A,\beta)<\frac{1}{2}
\Rightarrow
\gamma_s\left(\A,\frac{1}{1-\hat{\gamma}}\beta\right)
\leq \frac{\hat{\gamma}}{1-\hat{\gamma}}<1.\end{eqnarray}

To conclude the proof, we need to prove that
the inequalities we established:
\[
\hat{\gamma}_s \left(\A,\frac{1}{1+\hat{\gamma}}\beta\right)
\leq \frac{\gamma}{1+\gamma}
\mbox{ and }
\gamma_s \left(\A,\frac{1}{1-\hat{\gamma}}\beta\right)
\leq \frac{\hat{\gamma}}{1+\hat{\gamma}}
\]
are both equations.
This is straightforward by an argument
similar to the one in
the proof of Theorem 1 in \cite{JN08}. We omit it for
the sake of brevity.
{\qed}

We end this section by giving an equivalent representation
of the ${G}$-number $\hat{\gamma}_s(\A,\beta)$.
The next result generalizes Theorem 2 of \cite{JN08} (and its proof).
We define a compact convex set first:
\[
P_s:=\{Z\in \R^{m\times n}: \|Z\|_\ast\leq s, \|Z\|\leq 1\}.
\]
\begin{Theorem} \label{ft4}
Let $\A$ be a linear transformation, $\beta \in [0, +\infty]$ and
$s \in \{0,1, \ldots, r\}$.  Also let $P_s$ be as defined above.
Then,
\begin{eqnarray}\label{2f13}
\hat{\gamma}(\A,\beta)=\max_{Z,X}\{\langle Z,X\rangle-\beta\|\A X\|:
Z\in P_s, \|X\|_\ast\leq 1\}.\end{eqnarray} Moreover,
\begin{eqnarray}\label{2f14}
\hat{\gamma}(\A)=\max_{Z,X}\{\langle Z,X\rangle: Z\in P_s,
\|X\|_\ast\leq 1, \A X=0\}.\end{eqnarray}
\end{Theorem}
\textbf{Proof.} Let $B_\beta:= \{y\in\R^p: \|y\|_d\leq \beta\}$ and
$B:=\{X\in\R^{m\times n}: \| X\|\leq 1\}$. By definition,
$\hat{\gamma}_s(\A,\beta)$ is the smallest $\gamma$ such that the
closed convex set $C_{\gamma, \beta}:=\A^\ast B_\beta +\gamma B$
contains all matrices with $s$ nonzero singular values,
all equal to 1.
Equivalently,  $C_{\gamma,\beta}$ contains the convex hull of these
matrices, namely, $P_s$. Note that
$\gamma$ satisfies the
inclusion $P_s\subseteq C_{\gamma,\beta}$ if and only if
for every $X \in \R^{m\times n}$,
\begin{eqnarray}\max_{Z\in P_s}\langle Z,X\rangle\leq \max_{Y\in
C_{\gamma,\beta}}\langle Y,X\rangle &=&\max_{y\in \R^p,W\in
\R^{m\times n}}\{\langle X, \A^\ast y\rangle+\gamma \langle
X,W\rangle: \|y\|_d\leq \beta, \|W\|\leq 1\}\nonumber\\
\label{2f15}&=&\beta \|\A X\|+\gamma\|X\|_\ast.\end{eqnarray}
For the above, we adopt
the convention that whenever $\beta =+\infty$, $\beta\|\A X\|$  is
defined to be
$+\infty$ or $0$ depending on whether $\|\A X\| > 0$ or $\|\A X\| =
0$. Thus, $P_s\subseteq C_{\gamma,\beta}$ if and only if $\max_{Z\in
P_s}\{\langle Z,X\rangle-\beta \|\A X\|\}\leq \gamma\|X\|_\ast$.
Using the
homogeneity of this last relation
with respect to $X$, the above is equivalent to
$$\max_{Z,X}\{\langle Z,X\rangle-\beta \|\A X\|:Z\in P_s,
\|X\|_\ast\leq 1 \}\leq \gamma.$$ Therefore, the desired conclusion
holds. {\qed}

%\section{Characterization of $s$-goodness via $G$-numbers}
\section{$S$-goodness and $G$-numbers}
%\subsection{Exact Solution verification Quantity $\gamma_s(\A)$ and $s$-goodness}
We first give the following characterization result of $s$-goodness of
a linear transformation $\A$ via
the $G$-number $\gamma_s(\A)$, which
explains the importance of $\gamma_s(\A)$  in LMR. In the case of
SSR, it reduces to Theorem 1(i) in \cite{JN08}.

\begin{Theorem}\label{ft1} Let  $\A:\R^{m\times n}\rightarrow \R^p$ be a linear
transformation, and $s$ be an integer $s\in\{0,1,2,\ldots,r\}$. Then $\A$ is $s$-good
 if and only if $\gamma_s(\A)<1$.
\end{Theorem}
\textbf{Proof.} Suppose $\A$ is $s$-good.
Let $W\in\R^{m\times n}$ be a matrix of rank
$s\in\{1,2,\ldots,r\}$. Without loss of generality, let $W=U_{m\times
s}W_sV^T_{n\times s}$ be its SVD where
$U_{m\times s}\in\R^{m\times s}, V_{n\times s}\in\R^{n\times s}$ are
orthogonal matrices and
$W_s=\Diag((\sigma_1(W),\ldots,\sigma_s(W))^T)$. By the definition of
$s$-goodness of $\A$, $W$ is the unique solution to the optimization
problem (\ref{2f2}). Using the first order
optimality conditions, we obtain
that there
exists $y\in \R^p$ such that
the function $f_y(x)=\|X\|_\ast-y^T[\A
X-\A W]$ attains its minimum value
over $X\in \R^{m\times n}$ at $X=W$.
So, $0\in
\partial f_y(W)$, or $\A^\ast y\in\partial \|W\|_\ast$. Using the
fact (see, e.g., \cite{W92}) $$\partial \|W\|_\ast=\{U_{m\times
s}V^T_{n\times s}+M: W ~{\rm and}~ M {\rm~ have
~orthogonal~row~and~column~spaces,~and}~\|M\|\leq 1\},$$ it
follows that there exist matrices $U_{m\times (r-s)}, V_{n\times
(r-s)}$ such that $\A^\ast y=U \Diag(\sigma_i(\A^\ast y))V^T$ where
$U=[U_{m\times s}~U_{m\times (r-s)}]$, $V=[V_{n\times s}~V_{n\times
(r-s)}]$ are orthogonal matrices and
$$\sigma_i(\A^\ast y) \left \{\begin{array}{cc} =1,  &
\textrm{if}~~i\in J,\cr\noalign{\vskip2truemm}
\in[0,1],  & \textrm{if}~~i\in \bar{J}, \\
\end{array}\right.$$
where $J:=\{i:\sigma_i(W)\neq 0\}$ and
$\bar{J}:=\{1,2,\ldots,r\}\setminus J$. Therefore, the
optimal objective value of the
optimization problem
\begin{eqnarray}\label{2f7}
\min_{y,\gamma}\left\{\gamma:\A^\ast y\in\partial \|W\|_\ast, \sigma_i(\A^\ast y) \left
\{\begin{array}{cc} =1,  & \textrm{if}~~i\in
J,\cr\noalign{\vskip2truemm}
\in[0,\gamma],  & \textrm{if}~~i\in \bar{J}, \\
\end{array}\right.\right\}
\end{eqnarray}
is at most one. For the given $W$ with its SVD $W=U_{m\times
s}W_sV^T_{n\times s}$, let
$$\Pi:={\rm conv}\{M\in\R^{m\times n}:
{\rm ~the ~SVD~ of } M {\rm ~is~}
M=[U_{m\times s}~\bar{U}_{m\times (r-s)}]\left(
\begin{array}{cc}
0_s & 0 \\
0 & \sigma(M)\\
\end{array}
\right)
[V_{n\times s}~\bar{V}_{n\times
(r-s)}]^T\}.$$
It is easy to see that $\Pi$ is a subspace and
its normal cone (in the sense of variational analysis, see,
e.g., \cite{RW04} for details) is specified by
$\Pi^\perp.$
Thus, the above problem (\ref{2f7}) is equivalent to the following convex optimization problem with set constraint
\begin{eqnarray}\label{2f8}
\min_{y,M}\left\{\|M\|:\A^\ast y-U_{m\times
s}V^T_{n\times s}-M=0, M\in \Pi\right\}.
\end{eqnarray}
We will show that the optimal value is less than $1$.  For a contradiction,
suppose that
the optimal value is one.  Then, by Theorem 10.1 and Exercise 10.52 in \cite{RW04}, there exist
Lagrange multiplier
$D\in\R^{m\times n}$ such that the function
$$L(y,M)=\|M\|+\langle D, \A^\ast y-U_{m\times
s}V^T_{n\times s}-M\rangle+\delta_{\Pi}(M)$$ has unconstrained
minimum in $y, M$ equal to $1$, where $\delta_{\Pi}(\cdot)$ is the indicator
function of $\Pi$. Let $y^\ast, M^\ast$ be an optimal solution.
Then, by the optimality condition $0\in  \partial L$, we obtain that $$0\in \partial_y L(y^\ast, M^\ast), ~{\rm and} ~0\in \partial_M L(y^\ast, M^\ast).$$
Direct calculation yields that $$\A D=0, ~{\rm and}~ 0\in -D+\partial\|M^\ast\|+{\Pi}^\perp.$$ Notice that Corollary 6.4 in \cite{LS05} implies that for
every $C\in\partial\|M^\ast\|$, $C\in \Pi$ and $\|C\|_\ast\leq 1$. Then there exist $D_J\in {\Pi}^\perp $ and $D_{\bar{J}}\in\partial\|M^\ast\|\subset\Pi$ such that $D=D_J+D_{\bar{J}}$ with $\|D_{\bar{J}}\|_\ast\leq 1.$ Therefore, $\langle D, U_{m\times
s}V^T_{n\times s}\rangle=\langle D_J, U_{m\times
s}V^T_{n\times s}\rangle$ and  $\langle D, M^\ast\rangle=\langle D_{\bar{J}},M^\ast\rangle$. Moreover,  $\langle D_{\bar{J}}, M^\ast\rangle\leq\|M^\ast\|$ by the definition of the dual norm of $\|\cdot\|$. This together with the facts
$\A D=0$, $D_J\in {\Pi}^\perp $ and
$D_{\bar{J}}\in\partial\|M^\ast\|\subset\Pi$ yields
\begin{eqnarray} L(y^\ast,M^\ast)&=&\|M^\ast\|-\langle D_{\bar{J}},M^\ast\rangle+\langle D, \A^\ast y^\ast\rangle-\langle D_J, U_{m\times
s}V^T_{n\times s}\rangle+\delta_{\Pi}(M^\ast)\nonumber\\
&\geq & -\langle D_J, U_{m\times
s}V^T_{n\times s}\rangle+\delta_{\Pi}(M^\ast).\nonumber\end{eqnarray}
Thus, the minimum value of  $L(y,M)$ is attained,
$L(y^\ast,M^\ast)=-\langle D_J, U_{m\times s}V^T_{n\times s}\rangle$, when $M^\ast\in\Pi, \langle D_{\bar{J}}, M^\ast\rangle=\|M^\ast\|$.
We obtain that   $\|D_{\bar{J}}\|_\ast=1$. By assumption,
$1=L(y^\ast,M^\ast)=-\langle D_J, U_{m\times s}V^T_{n\times s}\rangle$. That is, $\sum_{i=1}^s(U_{m\times s}^TDV_{n\times s})_{ii}=-1.$
Without loss of generality, let SVD of the optimal $M^\ast$ be
$M^\ast=\tilde{U}\left(
\begin{array}{cc}
0_s & 0 \\
0 & \sigma(M^\ast)\\
\end{array}
\right)
\tilde{V}^T$, where $\tilde{U}
:=[U_{m\times s}~\tilde{U}_{m\times (r-s)}]$
and $\tilde{V}:=[V_{n\times s}~\tilde{V}_{n\times (r-s)}]$.
From the above arguments, we obtain that

   i) $\A D=0,$

  ii) $\sum_{i=1}^s(U_{m\times s}^TDV_{n\times s})_{ii}
=\sum_{i\in J}(\tilde{U}^TD\tilde{V})_{ii}=-1,$

  iii) $\sum_{i\in \bar{J}}(\tilde{U}^TD\tilde{V})_{ii}=1.$

\noindent Clearly, for every $t \in \R$, the matrices $X_t:=W+tD$ are
feasible in (\ref{2f2}). Note that $$W=U_{m\times
s}W_sV^T_{n\times s}=[U_{m\times s}~\tilde{U}_{m\times (r-s)}]\left(
\begin{array}{cc}
W_s & 0 \\
0 & 0\\
\end{array}
\right)
[V_{n\times s}~\tilde{V}_{n\times (r-s)}]^T.$$
Then, $\|W\|_\ast=\|\tilde{U}^TW\tilde{V}\|_\ast={\trace}(\tilde{U}^TW\tilde{V})$. From the above equations, we obtain
that $\|X_t\|_\ast=\|W\|_\ast$ for all small enough $t>0$
(since $\sigma_i(W) >0$, $i \in \{1,2, \ldots, s\}$). Noting that
$W$ is the unique optimal solution to (\ref{2f2}), we have $X_t=W$,
which means that $(\tilde{U}^TD\tilde{V})_{ii}=0$
for $i\in J$. This is a contradiction,
and hence the desired conclusion holds.

We next prove that $\A$ is $s$-good if $\gamma_s(\A)<1$. That is,
we let $W$ be an $s$-rank matrix and we show that $W$ is the
unique optimal solution to (\ref{2f2}). Without loss of generality,
let $W$ be a matrix of rank $s'\neq 0$ and $U_{m\times
s'}W_{s'}V^T_{n\times s'}$ be its SVD,
where $U_{m\times s'}\in\R^{m\times s'}, V_{n\times
s'}\in\R^{n\times s'}$ are orthogonal matrices and
$W_{s'}=\Diag((\sigma_1(W),\ldots,\sigma_{s'}(W))^T)$. It follows from
Proposition \ref{ft02} that $\gamma_{s'}(\A)\leq \gamma_s<1$. By the
definition of $\gamma_s(\A)$, there exists $y\in \R^p$ such that
$\A^\ast y=U\Diag(\sigma(\A^\ast y))V^T$, where $U=[U_{m\times
s'}~U_{m\times (r-s')}]$, $V=[V_{n\times s'}~V_{n\times (r-s')}]$
$$\sigma_i(\A^\ast y) \left \{\begin{array}{cc} =1,  &
\textrm{if}~~\sigma_i(W)\neq 0,\cr\noalign{\vskip2truemm}
\in[0,1),  & \textrm{if}~~\sigma_i(W)=0. \\
\end{array}\right.$$
The function $$f(X)=\|X\|_\ast-y^T[\A X-\A W]=\|X\|_\ast-\langle
\A^\ast y, X\rangle+\|W\|_\ast$$ becomes the objective function of
(\ref{2f2}) on the feasible set of (\ref{2f2}). Note that $\langle
\A^\ast y, X\rangle\leq \|X\|_\ast$ by $\|\A^\ast y\|\leq 1$ and the
definition of dual norm. So,
$f(X)\geq\|X\|_\ast-\|X\|_\ast+\|W\|_\ast=\|W\|_\ast$ and this
function attains its unconstrained minimum in $X$ at $X=W$. Hence
$X=W$ is an optimal solution to (\ref{2f2}). It remains to show that
this optimal solution is unique. Let $Z$ be another optimal solution
to the problem. Then $f(Z)-f(W)=\|Z\|_\ast-y^T \A
Z=\|Z\|_\ast-\langle\A^\ast y,Z\rangle=0.$ This together with the
fact $\|\A^\ast y\|\leq 1$ imply that there exist SVDs
for $\A^\ast y$ and $Z$ such that:
$$\A^\ast y=\tilde{U}\Diag(\sigma(\A^\ast
y))\tilde{V}^T,~~Z=\tilde{U}\Diag(\sigma(Z))\tilde{V}^T,$$ where
$\tilde{U}\in\R^{m\times r}$ and $\tilde{V}\in\R^{n\times r}$ are
orthogonal matrices, and $\sigma_i(Z)=0$ if $\sigma_i(\A^\ast
y)\neq 1$. Thus, for $\sigma_i(\A^\ast y)=0,
\forall i\in\{s'+1,\ldots,r\}$, we must have $\sigma_i(Z)=\sigma_i(W)=0$.
By the two forms of SVDs of $\A^\ast y$ as
above, $U_{m\times s'}V^T_{n\times s'}=\tilde{U}_{m\times
s'}\tilde{V}^T_{n\times s'}$ where $\tilde{U}_{m\times
s'},\tilde{V}^T_{n\times s'}$ are the corresponding submatrices of
$\tilde{U},\tilde{V}$, respectively. Without loss of generality, let
$$U=[u_1,u_2,\ldots,u_r],~V=[v_1,v_2,\ldots,v_r]~{\rm~ and~
}~~\tilde{U}=[\tilde{u}_1,\tilde{u}_2,\ldots,\tilde{u}_r],~
\tilde{V}=[\tilde{v}_1,\tilde{v}_2,\ldots,\tilde{v}_r],$$ where
$u_j=\tilde{u}_j$ and $v_j=\tilde{v}_j$ for the corresponding index
$j\in\{i:\sigma_i(\A^\ast y)=0 , i\in\{s'+1,\ldots,r\}\}$. Then we
have
$$Z=\sum_{i=1}^{s'}\sigma_i(Z)\tilde{u}_i\tilde{v}^T_i,~~W=\sum_{i=1}^{s'}\sigma_i(W){u}_i{v}^T_i.$$
From $U_{m\times s'}V^T_{n\times s'}=\tilde{U}_{m\times
s'}\tilde{V}^T_{n\times s'}$, we obtain that
$$\sum_{i=s'+1}^{r}\sigma_i(\A^\ast y)\tilde{u}_i\tilde{v}^T_i=\sum_{i=s'+1}^{r}\sigma_i(\A^\ast y){u}_i{v}^T_i.$$
Therefore, we deduce
\begin{eqnarray}&~&\sum_{i=s'+1,\sigma_i(\A^\ast y)\neq 0}^{r}\sigma_i(\A^\ast y)\tilde{u}_i\tilde{v}^T_i+\sum_{i=s'+1,\sigma_i(\A^\ast y)=0}^{r}\tilde{u}_i\tilde{v}^T_i
\nonumber\\
&=&\sum_{i=s'+1,\sigma_i(\A^\ast y)\neq 0}^{r}\sigma_i(\A^\ast
y){u}_i{v}^T_i+\sum_{i=s'+1,\sigma_i(\A^\ast
y)=0}^{r}{u}_i{v}^T_i\nonumber\\
&=:&\Omega.\nonumber\end{eqnarray}Clearly, the rank of $\Omega$ is
no less than $r-s'\geq r-s$. From the orthogonality
property of $U,V$ and
$\tilde{U},\tilde{V}$, we easily derive that
$$\Omega^T\tilde{u}_i\tilde{v}^T_i=0,~~\Omega^T{u}_i{v}^T_i=0, {\rm ~~for~all}~i\in\{1,2,\ldots,s'\}.$$
Thus, we obtain $\Omega^T(Z-W)=0$, which implies that the rank of
the matrix $Z-W$ is no more than $s$. Since $\gamma_s(\A)<1$, there
exists $\tilde{y}$ such that
$$\sigma_i(\A^\ast \tilde{y}) \left \{\begin{array}{cc} =1,  &
\textrm{if}~~\sigma_i(Z-W)\neq 0,\cr\noalign{\vskip2truemm}
\in[0,1),  & \textrm{if}~~\sigma_i(Z-W)=0. \\
\end{array}\right.$$ Therefore,
$0=\tilde{y}^T\A(Z-W)=\langle\A^\ast\tilde{y},Z-W\rangle=\|Z-W\|_\ast.$
Then $Z=W$.{\qed}

For the ${G}$-number $\hat{\gamma}_s(\A)$,
we directly obtain the following equivalent theorem of
$s$-goodness from Proposition \ref{ft3} and Theorem \ref{ft1}.

\begin{Theorem}\label{ft2} Let  $\A:\R^{m\times n}\rightarrow \R^p$ be a linear
transformation, and $s\in\{1,2,\ldots,r\}$.
Then $\A$ is $s$-good if and only if $\hat{\gamma}_s(\A) < 1/2.$
\end{Theorem}

For $X\in\R^{m\times n}$, we define the sum of the $s$ largest
singular values of $X$ as $$\|X\|_{s,\ast} :=\max_{Z\in P_s}\langle
Z,X\rangle.$$ We immediately obtain the following result utilizing
Proposition \ref{ft3} and Theorem \ref{ft2}.

\begin{Corollary} \label{ft5} Let $\A:\R^{m\times n}\rightarrow \R^p$ be
a linear
transformation,  and $s\in\{1,2,\ldots,r\}$. Then
$\hat{\gamma}_s(\A)$ is the best upper bound on the norm
$\|X\|_{s,\ast}$ of matrices $X\in {\rm Null}(\A)$ such that
$\|X\|_\ast\leq 1.$ As a result, the
linear transformation $\A$ is
$s$-good if and only if the maximum of $\|\cdot\|_{s,\ast}$-norms of
matrices $X\in{\rm Null}(\A)$ with $\|X\|_\ast= 1$ is less than $1/2$.\end{Corollary}

%\section{Error bounds for imperfect low-rank matrix recovery}
 \section{Exact and stable recovery via $G$-number}

In the previous sections, we showed that $G$-numbers $\gamma_s(\A)$
and $\hat{\gamma}_s(\A)$ are responsible for $s$-goodness of a linear
transformation $\A$. Observe that the definition of
$s$-goodness of a linear transformation $\A$
indicates that whenever the observation $b$ in
the following
\begin{eqnarray}\label{2f1} \hat{W}\in
{\rm argmin}_X\{\|X\|_\ast:\|\A X-b\|\leq\varepsilon\}\end{eqnarray}
is exact (noiseless) and comes from a $s$-rank matrix $W$ such
that $b=\A W$, $W$
is the unique optimal solution of the above optimization
problem (\ref{2f1}) where $\varepsilon$ is set to $0$. This establishes
a sufficient condition for the precise LMR of
an $s$-rank matrix $W$ in the ``ideal case" when there is no
measurement error or noise and the optimization problem (\ref{2f2}) is solved
exactly.

\begin{Theorem} \label{ft6-0} Let  $\A:\R^{m\times n}\rightarrow \R^p$ be a linear
transformation, and $s\in\{1,2,\ldots,r\}$.
Let $W$ be a $s$-rank matrix such that
$\A W=b$. If $\A $ is $s$-good ($\hat{\gamma}_s(\A)<1/2$, or $\gamma_s(\A)<1$), then $W$ is the unique solution to LMR (\ref{1f1}), i.e., the solution to LMR (\ref{1f1}) can be exactly recovered from Problem (\ref{2f2}).\end{Theorem}
\textbf{Proof.} By the definition of $s$-goodness of a linear transformation $\A$, the assumptions that $\A W=b$ and rank$(W)\leq s$ imply
that $W$ is the unique solution to problem (\ref{2f2}).
It remains to show that $W$ is the unique solution to problem (\ref{1f1}).
For a contradiction, suppose there is an another solution $Y$ to problem (\ref{1f1}). Then $\A W=\A Y=b$. By the $s$-goodness of $\A$, the problem $\min\{\|X\|_\ast: \A X=\A W\}\thickapprox\min\{\|X\|_\ast: \A X=\A Y\}$ has a unique solution, hence $Y=W$ and we reached a contradiction.{\qed}

It turns out that the same quantities $\gamma_s(\A)$
($\hat{\gamma}_s(\A)$) can be used to measure the
error of low-rank matrix recovery in the case when the matrix
$W\in\R^{m\times n}$ is not $s$-rank and the problem (\ref{2f2}) is
not solved exactly. In what follows, let
$W=U\Diag(\sigma(W))V^T$, where
$\sigma(W)=(\sigma_1(W),\ldots,\sigma_r(W))^T$ and
$\sigma_1(W)\geq\ldots\geq\sigma_r(W)\geq 0$ are the singular
values of $W$ in nonincreasing order. Let
$W^s:=U\Diag((\sigma_1(W),\ldots,\sigma_s(W),0,\ldots,0)^T)V^T$.
Clearly, in terms of nuclear norm, $W^s$ stands for the best $s$-rank
approximation of $W$. In order to establish the error bound in the
``non-ideal case", we also need the following assumption for a
matrix $X\in\R^{m\times n}$:

\vskip3mm\textbf{Block Assumption}: We say that $X$
satisfies the \emph{block assumption} with respect to $W$ if there exists $(U,V)\in \Xi(W)$ such that $U^TXV$ has
the block form as
$$U^TXV=\left(
           \begin{array}{cc}
             X_1 &0 \\
             0 & X_2 \\
           \end{array}
         \right)
,$$ where $X_1\in\R^{s\times s}$ and $X_2\in\R^{(r-s)\times (r-s)}$.
In this case, we write $X^{(s)}:=U\left(
           \begin{array}{cc}
             X_1 &0 \\
             0 & 0 \\
           \end{array}
         \right)
V^T$.

\begin{Theorem} \label{ft6} Let  $\A:\R^{m\times n}\rightarrow \R^p$ be a linear
transformation,  $s\in\{0,1,2,\ldots,r\}$, and
$\hat{\gamma}_s(\A)<1/2$ (or, equivalently, $\gamma_s(\A)<1$).
Also let $W$ be a matrix such that
$\A W=b$. Let $X$ be a $\upsilon$-optimal solution to the
problem (\ref{2f2}), meaning that $$\A X = \A W~~{\rm
and}~~\|X\|_\ast\leq Opt(\A W) + \upsilon,$$ where Opt$(\A W)$ is
the optimal value of (\ref{2f2}). If  the Block Assumption holds for
$X$, then
$$\|X-W\|_\ast\leq\frac{\upsilon+2\|W-W^s\|_\ast}{1-2\hat{\gamma}_s(\A)}=\frac{1+{\gamma}_s(\A)}{1-{\gamma}_s(\A)}[\upsilon+2\|W-W^s\|_\ast].$$
\end{Theorem}
\textbf{Proof.} Set $Z:=X-W$. Let
$D_1:=\Diag((\sigma_1(W),\ldots,\sigma_s(W))^T), D_2:=
\Diag((\sigma_{s+1}(W),\ldots,\sigma_r(W))^T)$. Using the assumptions,
we obtain that $Z$ has the form $$Z=U\left(
           \begin{array}{cc}
             X_1-D_1 &0 \\
             0 & X_2-D_2 \\
           \end{array}
         \right)
V^T.$$ Define $$Z^{(s)}:=U\left(
           \begin{array}{cc}
             X_1-D_1 &0 \\
             0 & 0 \\
           \end{array}
         \right)
V^T.$$ It is easy to verify that $Z^{(s)}=X^{(s)}-W^s$ and
$\|Z^{(s)}\|_\ast\leq \|Z\|_{s,\ast}$.  Along with the fact $\A Z=0$
and Corollary \ref{ft5}, this yields
\begin{eqnarray}\label{2f16}\|Z^{(s)}\|_\ast\leq
\|Z\|_{s,\ast}\leq\hat{\gamma}_s(\A)\|Z\|_{\ast}.\end{eqnarray} On
the other hand, $W$ is a feasible solution to (\ref{2f2}), so
Opt$(\A W) \leq\|W\|_{\ast}$. Thus, we have
\begin{eqnarray}\|W\|_{\ast}+\upsilon\geq\|W+Z\|_{\ast}&\geq&\|W^s+Z-Z^{(s)}\|_{\ast}-\|Z^{(s)}+W-W^s\|_{\ast}\nonumber\\
\label{2f17}&=&\|W^s\|_{\ast}+\|Z-Z^{(s)}\|_{\ast}-\|Z^{(s)}\|_{\ast}-\|W-W^s\|_{\ast},\end{eqnarray}
where the last equation follows
from the facts that
$W^s(Z-Z^{(s)})^T=0=(W-W^s)(Z^{(s)})^T$~and $
(W^s)^T(Z-Z^{(s)})=0=(W-W^s)^T Z^{(s)}$, and Lemma 2.3 in
\cite{RFP08}. This is equivalent to
\begin{eqnarray}\|Z-Z^{(s)}\|_{\ast}\leq \|Z^{(s)}\|_{\ast}+2\|W-W^s\|_{\ast}+\upsilon.\nonumber\end{eqnarray}
Therefore, we obtain
\begin{eqnarray}\|Z\|_{\ast}\leq\|Z^{(s)}\|_{\ast}+\|Z-Z^{(s)}\|_{\ast}&\leq&
2\|Z^{(s)}\|_{\ast}+2\|W-W^s\|_{\ast}+\upsilon\nonumber\\
&\leq&2\hat{\gamma}_s(\A)\|Z|_{\ast}+2\|W-W^s\|_{\ast}+\upsilon.\nonumber\end{eqnarray}
Since $\hat{\gamma}_s(\A)<1/2$, we reach the desired conclusion.{\qed}

Notice that the  above Block Assumption holds naturally in the SSR
(CMP) context. In general, we may have $$U^TXV=\left(
           \begin{array}{cc}
             X_1 & X_3 \\
             X_4 & X_2 \\
           \end{array}
         \right),$$ where either $X_3$ or $X_4$ is not zero. In this case, we have
$$Z=U\left(
           \begin{array}{cc}
             X_1-D_1 & X_3 \\
             X_4 & X_2-D_2 \\
           \end{array}
         \right)
V^T.$$ If we define $$Z^{(s)}:=U\left(
           \begin{array}{cc}
             X_1-D_1 &0 \\
             0 & 0 \\
           \end{array}
         \right)
V^T,$$ we cannot conclude (\ref{2f17}). If we define $$Z^{(s)}:=U\left(
           \begin{array}{cc}
             X_1-D_1 & X_3 \\
             X_4 & 0 \\
           \end{array}
         \right)
V^T,$$ we cannot conclude
$\|Z^{(s)}\|_\ast\leq \|Z\|_{s,\ast}$. It is not difficult to
give counterexamples to illustrate the above facts. Meanwhile, in the last two cases, the rank of
$Z^{(s)}$ may be greater than $s$. Thus the condition
$\hat{\gamma}_s(\A)<1/2$ is not sufficient, and hence we need more
strict restrictions
on the linear transformation $\A$.

Below, we consider approximate solutions $X$ to the problem
\begin{eqnarray}\label{2f18} Opt(b)=\min_{X\in\R^{m\times n}}\{\|X\|_\ast: \|\A X-b\|\leq\varepsilon\}\end{eqnarray}
where $\varepsilon\geq 0$ and $b=\A W+\zeta, ~~~\zeta\in \R^p$
with $\|\zeta\|\leq \varepsilon$. We will show that in the
``non-ideal case", when $W$ is ``nearly $s$-rank" and (\ref{2f18})
is solved to near-optimality, the error of the LMR via NNM can be
measured by
$\hat{\gamma}_s(\A, \beta)$ with a finite $\beta$.

\begin{Theorem} \label{ft7} Let  $\A:\R^{m\times n}\rightarrow \R^p$ be a linear
transformation, and $s\in\{1,2,\ldots,r\}$, and let $\beta\in
[0,+\infty]$ such that $\hat{\gamma}:=\hat{\gamma}_s(\A,\beta)<1/2$
(or ${\gamma}:={\gamma}_s(\A,\beta/(1-\hat{\gamma}))<1$). Let
$\varepsilon\geq 0$ and let $W$ and $b$ in (\ref{2f18})  be such
that $\|\A W-b\|\leq \varepsilon$, and let $W^s$ be defined in the
beginning of this section. Let $X$ be a
$(\vartheta,\upsilon)$-optimal solution to the problem
(\ref{2f18}), meaning that $$\|\A X -b\|\leq\vartheta~~{\rm
and}~~\|X\|_\ast\leq Opt(b) + \upsilon.$$ If the Block Assumption
holds for $X$, then
\begin{eqnarray}\label{2f19}\|X-W\|_\ast&\leq&\frac{2\beta(\vartheta+\varepsilon)+2\|W-W^s\|_\ast+\upsilon}{1-2\hat{\gamma}}\nonumber\\
&=&\frac{1+{\gamma}}{1-{\gamma}}[2\beta(\vartheta+\varepsilon)+2\|W-W^s\|_\ast+\upsilon].\end{eqnarray}
\end{Theorem}
\textbf{Proof.} Note that $W$ is a feasible solution to
(\ref{2f18}). Let $Z=X-W$. As in the proof of Theorem \ref{ft6},
 we obtain that $\|Z^{(s)}\|_\ast\leq \|Z\|_{s,\ast}$ and
\begin{eqnarray}\|Z\|_{\ast}\leq
2\|Z^{(s)}\|_{\ast}+2\|W-W^s\|_{\ast}+\upsilon.\nonumber\end{eqnarray}
Employing (\ref{2f13}) in Theorem \ref{ft4}, we derive
\begin{eqnarray}\label{2f20}\|Z\|_{s,\ast}\leq\beta\|\A Z\|+\hat{\gamma}\|Z\|_{\ast}\leq
\beta(\vartheta+\varepsilon)+\hat{\gamma}\|Z\|_{\ast},\end{eqnarray}
where the last inequality holds by $\|\A Z\|=\|\A X-b+b-\A
Z\|\leq\|\A X-b\|+\|b-\A Z\|$. Combining with the above
inequalities, we obtain
$$\|Z\|_\ast\leq 2\beta(\vartheta+\varepsilon)+2\hat{\gamma}\|Z\|_\ast+2\|W-W^s\|_\ast+\upsilon.$$
Now, the desired conclusion
follows from the assumption $\hat{\gamma}<1/2$
and ${\gamma}=\hat{\gamma}/(1+\hat{\gamma})$. {\qed}

Theorem \ref{ft7} shows that under the Block Assumption the error
bound (\ref{2f19}) for imperfect low-rank matrix recovery can be
bounded in terms of $\hat{\gamma}_s (\A, \beta),\beta$, measurement
error $\varepsilon$, ``s-tail" $\|W-W^s\|_\ast$ and the accuracy
$(\vartheta, \upsilon)$ to which the estimate solves the program
(\ref{2f18}).  Note that we need
${\gamma}_s (\A, \beta)<1$ (or $\hat{\gamma}_s (\A, \beta)<1/2)$. However, the ``true" necessary and
sufficient condition for $s$-goodness is ${\gamma}_s (\A)<1$ (or $\hat{\gamma}_s (\A)<1/2)$.
Also, note that ${\gamma}_s (\A, \beta)= {\gamma}_s (\A)$
for all finite ``large enough" values of $\beta$, see Proposition
\ref{fp1} for details.

%\section{Computable upper bounds of $\hat{\gamma}$}
\section{Computing bounds on the ${G}$-number via convex optimization}
We showed that ${G}$-number
$\hat{\gamma}_s(\A, \beta)$ controls some of the
fundamental properties of a linear
transformation $\A$ relative to LMR.
Since it seems difficult to evaluate these
quantities exactly, we  will provide ways of computing
upper and lower bounds on these quantities $\hat{\gamma}_s(\A,
\beta)$ via convex optimization techniques.

\subsection{Computing lower bounds on $\hat{\gamma}_s(\A,\beta)$}

Note that $\hat{\gamma}_s(\A,\beta)\geq \hat{\gamma}_s(\A)$ for any
$\beta>0$ by Proposition \ref{ft01}. Therefore, we may establish a
lower bound for $\hat{G}$-numbers $\hat{\gamma}_s(\A,\beta)$ by giving such a bound
for $\hat{\gamma}_s(\A)$. We can bound $\hat{\gamma}_s(\A)$ from
below utilizing Theorem \ref{ft4}.
Recall von Neumann's trace inequality \cite{von37}:
$\langle Y,Z\rangle\leq \langle\sigma (Y),\sigma(Z)\rangle$
for every pair of matrices $Y,Z\in\R^{m\times n}$,
where the equality holds when $Y,Z$ share the same orthogonal
row and column spaces. In what follows, we define
$$\Xi(\A):=\{(U,V):  U\in \R^{m\times r},
V\in \R^{n\times r},\exists W=U\Diag(\sigma(W))V^T, \A W=0\}.$$
From the representation
(\ref{2f14}), we obtain
$$\hat{\gamma}(\A)=\max_{\Sigma\in P_s}f(\Sigma),
~~f(\Sigma)=\max_{X}\{\langle \Sigma,X\rangle: \|X\|_\ast\leq 1, \A X=0\}.$$
It is easy to see that $f(\Sigma)$ is convex.  Then, we solve
the convex optimization problem
\begin{eqnarray}\label{2f21}X_\Sigma\in {\rm argmax}_X \{ \langle \Sigma,X\rangle: \|X\|_\ast\leq 1, \A
X=0\},\end{eqnarray} we obtain a linear form $\langle
X_\Sigma,\Theta\rangle$ of $\Theta\in P_s$ which under-estimates
$f(\Theta)$ everywhere and agrees with $f(\Theta)$ when
$\Theta=\Sigma$. Notice  that
\begin{eqnarray}& &{\rm max}_X \{ \langle \Sigma,X\rangle: \|X\|_\ast\leq 1, \A
X=0\}\nonumber\\ &\geq&{\rm max}_{X,(U,V)\in\Xi(\A)} \{ \langle \Sigma,X\rangle: \|X\|_\ast\leq 1, \A
X=0, \Sigma=U \Diag(t)V^T, X_\Sigma=U \Diag(x_t)V^T\}.\nonumber
\end{eqnarray} Since we need only to focus the lower bound
via the above problem (\ref{2f21}), in this sense, we may set
 $\Sigma=U \Diag(t)V^T$ by choosing $(U,V)\in\Xi(\A)$ and $t\in\R^r$ with $ \|t\|_1\leq s,
\|t\|_\infty\leq 1$. Thus, we may obtain a lower bound from the following optimization problem:
$${\rm max}_{x_t}\{ \langle t,x_t\rangle: \|x_t\|_1\leq 1, \A
[U \Diag(x_t)V^T]=0\}.$$
For simplicity, we define $\A$ by
a set of $p$
matrices $A_i\in \R^{m\times n}, i \in \{1,2,\ldots,p\}$:
$$\A (\cdot)=(\langle A_1,\cdot\rangle,\langle A_2,\cdot\rangle,\ldots,\langle
A_p,\cdot\rangle)^T.$$ Thus, we may rewrite
\begin{eqnarray}\label{2f22}\A X_\Sigma=A x_t\end{eqnarray} where
$A\in \R^{p\times r}$ with $A_{ij}=(U^TA_iV)_{jj}$. In this sense,
we may formulate the convex optimization
problem (\ref{2f21}) as the following
group of LP problems
\begin{eqnarray}\label{2f23}x_t\in {\rm argmax}_x \{ \langle t,x\rangle: \|x\|_1\leq 1, A
x=0\}.\end{eqnarray}
The optimal solutions may not be unique because for a given
$\Sigma$ orthogonal matrices $U\in\R^{m\times r}, V\in\R^{n\times
r}$ are usually not unique. In order to establish a lower bound for
$\hat{\gamma}_s(\A)$, we may choose one pair $(U,V)\in\Xi(\A)$ and then solve
the corresponding LP (\ref{2f23}). We obtain a linear form $v^Tx_t$
of $v\in \Delta_s$ where $$\Delta_s:=\{x\in \R^{r}: \|x\|_1\leq s,
\|x\|_\infty\leq 1\}.$$ Therefore, we obtain a lower bound result on $\hat{\gamma}_s(\A)$ as follows:

\begin{Proposition}\label{fp2-5} Let $\A$ be specified as above
and $x_t$ given by (\ref{2f23}). Then,
$\max_{v\in \Delta_s}v^Tx_t$ is a lower bound on $\hat{\gamma}_s(\A)$.
\end{Proposition}

Clearly, the above bound is easily computable.
As in \cite{JN08}, we can use the standard sequential convex
approximation scheme for maximizing the convex function $f(\cdot)$
over $P_s$. In particular, we can run the iterative process
$$t_{k+1}\in {\rm argmax}_{v\in
\Delta_s}v^Tx_{t_k}, ~~~t_1\in \Delta_s, ~~~U \Diag(t_1)V^T\in P_s.$$
This leads to a monotone nondecreasing sequence of lower bounds
$t_k^Tx_{t_k}$ on $\hat{\gamma}_s(\A)$. We may
choose to terminate this
iterative process when the improvement in the
bounds falls below a given
tolerance, and we can start several runs from randomly chosen
points $t_1$ and orthogonal matrices $(U,V)\in\Xi(\A)$.

\subsection{Computing upper bounds on $\hat{\gamma}_s(\A,\beta)$}

For an arbitrary linear
transformation $\B$, we have
\begin{eqnarray}&~&\max_{\Sigma, X}\{\langle \Sigma,X\rangle: \|X\|_\ast\leq 1, \A X=0, \Sigma\in P_s\}\nonumber\\
\label{2f24}&~&=\max_{\Sigma, X}\{\langle \Sigma,X-\B^{\ast}\A
X\rangle: \|X\|_\ast\leq 1, \A X=0, \Sigma\in P_s\}.\end{eqnarray}
In the same way
as in (\ref{2f22}), we define $\B$ by
a set of $p$ matrices $B_k\in \R^{m\times n},
k \in \{1,2,\ldots,p\}$ and $\B^\ast$ as
$$\B^{\ast} (u)=\sum_{k=1}^p u_k B_k,
~~u=(u_1,u_2,\ldots,u_p)^T\in\R^p.$$ For simplicity, suppose (\ref{2f22}) holds.
Using a similar analysis, we choose all $B_j$ (simultaneously
diagonalizable) such that they have the singular
value decompositions $B_k=U\Diag(y_k)V^T~(y_k\in \R^r)$ and then
rewrite (\ref{2f24}) as
\begin{eqnarray}&~&\max_{\Sigma, X}\{\langle \Sigma,X-\B^{\ast}\A X\rangle:
\|X\|_\ast\leq 1, \A X=0, \Sigma\in P_s\}\nonumber\\
\label{2f25}&~&=\max_{t, x, U, V}\{\langle t,x-B^T A x\rangle:
\|x\|_1\leq 1, A x=0, t\in \Delta_s\},\end{eqnarray} where
$B^T:=[y_1,y_2,\ldots,y_p]$. If we fix
$U,V$, the above problem is easy to solve
as it was done in \cite{JN08}.
In this case,\begin{eqnarray}&~&\max_{t, x}\{\langle t,x-B^T A
x\rangle:
\|x\|_1\leq 1, A x=0, t\in \Delta_s\}\nonumber\\
&~&\leq\max_{t, x}\{\langle t,x-B^T A x\rangle:
\|x\|_1\leq 1, t\in \Delta_s\}\nonumber\\
&~&=\max_{t, i\in\{1,\ldots,r\}}\{\langle t, (I-B^T
A)e_i\rangle:  t\in \Delta_s\}\nonumber\\
&~&=\max_{ i\in\{1,\ldots,r\}}\max_{t\in \Delta_s}\{\langle t,
(I-B^T A)e_i\rangle\}=\max_{ i\in\{1,\ldots,r\}}\|(I-B^T
A)e_i\|_{s,1},\end{eqnarray} where
$\|x\|_{s,1}$ is the sum of the $s$
largest magnitudes of entries in $x$. Therefore, we have for all
$B\in \R^{p\times r}$
\begin{eqnarray}\hat{\gamma}_s(\A)&=&\max_{\Sigma, X}\{\langle \Sigma,X\rangle: \|X\|_\ast\leq 1, \A X=0, \Sigma\in P_s\}\nonumber\\
&\leq &
\max_{U,V,i\in\{1,\ldots,r\}}\|(I-B^TA)e_i\|_{s,1}=:f_{\A,s}(B).\nonumber\end{eqnarray}
Taking $\Gamma_s(\A,+\infty):=\min_B f_{\A,s}(B)$, we obtain
$$\hat{\gamma}_s(\A)\leq \Gamma_s(\A,+\infty).$$
Observe that $f_{\A,s}(B)$ is an easy-to-compute convex function of
$B$ for fixed $U,V$ and it is indeed related to a semi-infinite
programming \cite{BS00}. Therefore, one may choose to
utilize computational semi-infinite
programming techniques to compute the quantity
$\Gamma_s(\A,+\infty)$.

The above analysis motivates the following useful function
of $\A$ and $\beta$.

\begin{Definition} Let $\A$ and the corresponding matrices $A_i, i\in\{1,2,\ldots,p\}$ be given as above.
Let $\beta\in [0,+\infty]$. We define  $\Gamma_s(\A,\beta)$ as follows:
\begin{eqnarray}\label{2f26}{\Gamma}_s(\A,\beta)&:=&\min_B
\left\{\max_{U,V, i\in\{1,\ldots,r\}}\|(I-B^T
A)e_i\|_{s,1}:\|(B)_{\cdot j}\|_d\leq\beta, 1\leq j\leq
r\right\},\end{eqnarray}
where  $A$ is the matrix defined by
$A_i$ and $U,V$ (as above), $(B)_{\cdot j}$ is the $j$th column of $B$.
If  there does not exist such a matrix $B$ as above, we take
$\Gamma_s(\A,\beta)=+\infty$. For convenience, we abbreviate
the notation
$\Gamma_s(\A,+\infty)$ to $\Gamma_s(\A)$.
\end{Definition}

By modifying the above process, we obtain that
$\Gamma_s(\A,\beta)$ provides an upper bound for
$\hat{G}$-numbers $\hat{\gamma}_s(\A,\beta)$.
Moreover, $\Gamma_s(\A,\beta)$ shares some
properties similar to those
of $\hat{G}$-numbers $\hat{\gamma}_s(\A,\beta)$. In other words,
$\Gamma_s(\A,\beta)$ is nondecreasing in $s$, convex and
nonincreasing in $\beta$, and is such that
$\Gamma_s(\A,\beta)=\Gamma_s(\A)$ for all large enough values of
$\beta$. The following result shows that $\Gamma_s(\A,\beta)$ is an
upper bound on $\hat{\gamma}_s(\A,\beta)$.

\begin{Theorem} \label{ft8}
For every $\A$ and $\beta \in [0, +\infty]$, we have
$\Gamma_s(\A,\beta)\geq\hat{\gamma}_s(\A,\beta)$.
\end{Theorem}
\textbf{Proof.} Let $W$ be a $s$-rank matrix with all
nonzero singular
values equal to $1$ such that $W=U\left(
                                    \begin{array}{cc}
                                      I_s & 0\\
                                      0& 0 \\
                                    \end{array}
                                  \right)
V^T,$ where $I_s$ is the $s\times s$ identity
matrix. For $U,V$, we get $\A
W=A\sigma(W)$ where $A$ is specified as in (\ref{2f22}) . Let
$Y=[y_1,y_2,\ldots, y_r]\in\R^{p\times r}$ be such that
$\|y_i\|_d\leq \beta$ and the columns in $I-Y^T A$ are of the
$\|\cdot\|_{s,1}$-norm not exceeding $\Gamma_s(\A,\beta)$. Define
the linear transformation $\B$ such that $\B W:=Y\sigma(W)$.
Setting $y=Y\sigma(W)$, the fact that $\|y_i\|_d\leq \beta$,
$i\in\{1,2,\ldots,r\}$ implies that
$\|y\|_\ast\leq\beta\|\sigma(W)\|_1\leq\beta s$. Furthermore,
noting that $\sigma(W)$ is a $s$-sparse vector, we obtain
$$\|W-\A^\ast y\|=\|W-\A^\ast \B W\|=\|(I-B^T
A)^T\sigma(W)\|\leq\Gamma_s(\A,\beta).$$The desired conclusion
follows immediately.  {\qed}

Note that $\|X\|_{st,\ast}\leq s\|X\|_{t,\ast}$ for all positive integers $s,t$.
Thus, we may replace  $\Gamma_s(\A,\beta)$ as $s{\Gamma}_1(\A,\beta)$, i.e.,
$$\hat{\gamma}_s(\A,\beta)\leq\Gamma_s(\A,\beta)\leq s{\Gamma}_1(\A,\beta).$$
Moreover, we have $\Gamma_1(\A,\beta)= \max_i \Upsilon_i$, where
\begin{eqnarray}\label{2f28}\Upsilon_i:=\min_{U,V,y_i}\{\|e_i-A^Ty_i\|_\infty: \|y_i\|_d\leq\beta\},~~i\in\{1,2,\ldots,r\}.\end{eqnarray}
By direct calculation, note that the matrix $A$ is the representation
of $\A$ with respect to $U,V$, we obtain
\begin{eqnarray}\Upsilon_i&=&\min_{U,V,y}\max_j\{|(e_i-A^Ty)_j|: \|y\|_d\leq\beta\}\nonumber\\
&=&\min_{U,V,y}\max_x\{\langle e_i-A^Ty, x\rangle: \|y\|_d\leq\beta, \|x\|_1\leq 1\}\nonumber\\
&=&\max_{X}\min_{y}\{\langle Ue_iV^T, X\rangle-\langle \A^\ast y, X\rangle: \|y\|_d\leq\beta, \|X\|_\ast\leq 1, X=U\Diag(x)V^T\}\nonumber\\
&=&\max_{X}\{\langle Ue_iV^T, X\rangle-\beta\|\A X\|: \|X\|_\ast\leq 1\}.\nonumber\end{eqnarray}
It follows from Theorem \ref{ft4} that $\Upsilon_i\leq \hat{\gamma}_1(\A,\beta)$. Therefore, by Theorem \ref{ft8}, we
have that the relaxation for $\hat{\gamma}_1(\A,\beta)$ is exact,
i.e.,
\begin{eqnarray}\label{2f29}
\Gamma_1(\A,\beta)=\hat{\gamma}_1(\A,\beta).\end{eqnarray}

As in Proposition \ref{fp1},
we present the following simple result which shows how large
$\beta$ needs to be to guarantee $\Gamma_s(\A,\beta)={\Gamma}_s(\A)$.

\begin{Proposition}\label{fp2}
Let  $\A:\R^{m\times n}\rightarrow \R^p$ be a linear
transformation, $\beta\in [0,+\infty]$ and
$s\in\{0,1,2,\ldots,r\}$. For some $\rho>0$, let
the image of the unit $\|\cdot\|_\ast$-ball in
$\R^{m\times n}$ under the mapping $X\mapsto\A X$ contain the ball $B=\{x\in\R^p: \|x\|_1\leq\rho\}$.
Then for every $s\leq r$
$$\beta\geq \frac{3}{2\rho}
\mbox{ and } \Gamma_s(\A)<\frac{1}{2}
~~\Rightarrow~~{\Gamma}_s (\A, \beta)= {\Gamma}_s (\A).$$
\end{Proposition}
\textbf{Proof.} Fix $s\in\{1,2,\ldots,r\}$. Let $\Gamma_s(\A)<1/2$. Then $\gamma:=\hat{\gamma}_s(\A)<1/2$ and hence
for every matrix $W\in \R^{m\times n}$ with $s$ nonzero
singular values, equal to $1$, there exists a vector $y\in\R^p$ such
that
\begin{eqnarray}\label{2p4} \|y\|_d\leq \beta {\rm~and~}
\|\A^\ast y-X\|\leq\gamma.
\nonumber\end{eqnarray}
By the triangle inequality,
$\|\A^\ast y\|\leq 1+\gamma<3/2$. Following the same steps
as in the proof of Proposition \ref{fp1},
we reach the desired conclusion.
{\qed}

\section{$S$-goodness and RIP}
We consider the connection between restricted isometry property
and $s$-goodness of
the linear transformation in LMR and
present some explicit forms of restricted isometry (RI)
constants and $s$-goodness constants, $G$-numbers.
Recall that  the
$s$-\emph{restricted isometry constant}
$\delta_s$ of a linear transformation $\A$ is defined as
the smallest constant such that the
following holds for all $s$-rank matrices
$X\in\R^{m\times n}$
\begin{eqnarray}\label{2f27}(1-\delta_s)\|X\|_F^2\leq \|\A X\|_2^2\leq (1+\delta_s)\|X\|_F^2.\end{eqnarray}
In this case, we say $\A$ possesses the \emph{RI$(\delta_s)$-property (RIP)} as in the CS context. For details, see \cite{RFP08,CP09,LB09,MJD09,MF10} and
the references therein.

\subsection{$\hat{\gamma}_s(\A)$ and $\delta_{2s}$}
We will show that the RI$(\delta_{2s}$)-property
of $\A$ implies that
$G$-numbers
satisfy $\hat{\gamma}_s(\A)<1/2$ and $\gamma_s(\A)<1$,
which means that the RIP implies the sufficient
conditions for $s$-goodness.

\begin{Theorem}\label{ft9}
Let  $\A:\R^{m\times n}\rightarrow \R^p$ be a linear transformation,
and $s\in\{1,2,\ldots,r\}$.  Assume that
$\A$ has RIP with $\delta_{2s}<\sqrt{2}-1$, and
let $\|\cdot\|_d:=\|\cdot\|_2$ for vectors in $\R^p$.
Then we have
\begin{eqnarray}\label{2f33} \hat{\gamma}_s(\A,\beta)\leq \frac{\sqrt{2}\delta_{2s}}{1+(\sqrt{2}-1)\delta_{2s}}<\frac{1}{2}
 {~~\rm~for~all}~~\beta\geq \frac{\sqrt{s(1+\delta_{2s})}}{1+(\sqrt{2}-1)\delta_{2s}}.\end{eqnarray}
This implies
\begin{eqnarray}\label{2f30} \hat{\gamma}_s(\A)\leq \frac{\sqrt{2}\delta_{2s}}{1+(\sqrt{2}-1)\delta_{2s}}<\frac{1}{2}
 {\rm~and~}{\gamma}_s(\A)\leq \frac{\sqrt{2}\delta_{2s}}{1-\delta_{2s}}<1,\end{eqnarray}
 and hence $\A$ is $s$-good.
\end{Theorem}
\textbf{Proof.} By Theorem \ref{ft4}, in order to show (\ref{2f33}), it is enough to verify that for all $X\in\R^{m\times n}$
 \begin{eqnarray}\label{2f34}\|X\|_{s,\ast}\leq \frac{\sqrt{s(1+\delta_{2s})}}{1+(\sqrt{2}-1)\delta_{2s}}\|\A X\|_2+\frac{\sqrt{2}\delta_{2s}}{1+(\sqrt{2}-1)\delta_{2s}}\|X\|_\ast.\end{eqnarray} Without loss of generality,
let SVD of $X$ be specified by $$X=U\Diag(x)V^T,$$
where $U\in
\R^{m\times r}$ and $V\in \R^{n\times r}$, and
$\sigma(X):=x=(x_1,\ldots,x_r)^T$
is the vector of the singular values of $X$ with
$x_1\geq\cdots\geq x_r\geq 0$. We
decompose $x$ into a sum of vectors
$x_{T_i}, i \in \{0,1,2,\ldots \}$,
each of sparsity at most $s$, where $T_0$ corresponds to the locations of the
$s$ largest entries of $X$, and $T_1$ to the locations
of the next $s$ largest entries, and so on (with except for the last part). We define
$X_{T_i}:=U\Diag(x_{T_i})V^T.$
Then, $X_{T_0}$ is the part of $X$ corresponding to the $s$
largest singular values, $X_{T_1}$ is the part corresponding to the
next $s$ largest singular values, and so on. Clearly,
$X_{T_0},X_{T_1},\ldots,X_{T_i},\ldots$ are all orthogonal
to one another,  and
rank($X_{T_i})\leq s$. From the above partition, we easily
obtain that for $j\geq 2$, $$\|X_{T_j}\|_F\leq
s^{1/2}\|X_{T_j}\|\leq s^{-1/2} \|X_{T_{j-1}}\|_\ast.$$
Then it follows that
\begin{eqnarray} \sum_{j\geq 2}\|X_{T_j}\|_F\leq s^{-1/2}\sum_{j\geq 2} \|X_{T_{j-1}}\|_\ast\leq s^{-1/2}(\|X\|_\ast-\|X_{T_0}\|_\ast).\nonumber\end{eqnarray}
This yields
\begin{eqnarray}\label{2f35}\|
X-X_{T_0}-X_{T_1}\|_F=\|\sum_{j\geq 2}X_{T_j}\|_F\leq\sum_{j\geq 2}\|X_{T_j}\|_F\leq s^{-1/2}(\|X\|_\ast-\|X_{T_0}\|_\ast).\end{eqnarray}
Noting that $\A
(X_{T_0}+X_{T_1})= \A
(X-\sum_{j\geq2}X_{T_j})$, we obtain
\begin{eqnarray}\|\A (X_{T_0}+X_{T_1})\|_2^2&=&\langle \A
(X_{T_0}+X_{T_1}), \A
(X-\sum_{j\geq2}X_{T_j})\rangle\nonumber\\&=&\langle \A
(X_{T_0}+X_{T_1}), \A
X\rangle-\sum_{j\geq2} \langle \A (X_{T_0}+X_{T_1}),   \A X_{T_j}\rangle.\nonumber\end{eqnarray}
From the RIP assumption of $\A$, we obtain that
\begin{eqnarray}|\langle \A (X_{T_0}+X_{T_1}), \A X\rangle|&\leq & \| \A (X_{T_0}+X_{T_1})\|_2 \|\A X\|_2
\nonumber\\
&\leq & \sqrt{1+\delta_{2s}}\|
X_{T_0}+X_{T_1}\|_F\|\A X\|_2.\nonumber\end{eqnarray}
By direct
calculation,
\begin{eqnarray}\sum_{j\geq2} |\langle \A
(X_{T_0}+X_{T_1}), \A X_{T_j})\rangle|&\leq &\sum_{j\geq2} \delta_{2s}(\|X_{T_0}\|_F+\|X_{T_1}\|_F) \|X_{T_j}\|_F\nonumber\\
&\leq &\sqrt{2}\delta_{2s}\|X_{T_0}+X_{T_1}\|_F\sum_{j\geq2}  \|X_{T_j}\|_F,\nonumber\end{eqnarray} where the first
inequality follows from Lemma 3.3 \cite{CP09}, and the second one follows from the inequality $(\|X_{T_0}\|_F+\|X_{T_1}\|_F)^2\leq 2\|X_{T_0}+X_{T_1}\|_F^2$.
Clearly, combining the  RIP assumption on
$\A$ with the above inequalities, we have
\begin{eqnarray}(1-\delta_{2s})\|X_{T_0}+X_{T_1}\|_F^2&\leq&\langle \A
(X_{T_0}+X_{T_1}), \A (X_{T_0}+X_{T_1})\rangle\nonumber\\
&\leq & \sqrt{1+\delta_{2s}}\|
X_{T_0}+X_{T_1}\|_F\|\A X\|_2+\sqrt{2}\delta_{2s}\|X_{T_0}+X_{T_1}\|_F\sum_{j\geq2}  \|X_{T_j}\|_F.\nonumber\end{eqnarray}
This implies
\begin{eqnarray}(1-\delta_{2s})\|X_{T_0}+X_{T_1}\|_F&\leq&\sqrt{1+\delta_{2s}}\|\A X\|_2+\sqrt{2}\delta_{2s}\sum_{j\geq2}  \|X_{T_j}\|_F.\nonumber\end{eqnarray}
By (\ref{2f35}) and the fact $\|X_{T_0}\|_\ast \leq \sqrt{s}\|X_{T_0}\|_F\leq \sqrt{s}\|X_{T_0}+X_{T_1}\|_F$, it follows that
$$\|X_{T_0}\|_\ast \leq \frac{\sqrt{s(1+\delta_{2s})}}{1-\delta_{2s}}\|\A X\|_2+\frac{\sqrt{2}\delta_{2s}}{1-\delta_{2s}}(\|X\|_\ast-\|X_{T_0}\|_\ast).$$
Noting that $\|X_{T_0}\|_\ast=\|X\|_{s,\ast}$, we
establish (\ref{2f34}), and hence we obtain the desired conclusion.
{\qed}

\subsection{$\Gamma_s(\A)$ and $\delta_{2s}$}
We consider the performance of $\Gamma_s(\A)$ for $s$-goodness when $\A$
has RIP. It turns out that this is similar to the CS case.
\begin{Theorem}\label{ft11} Let  $\A:\R^{m\times n}\rightarrow \R^p$ be a linear
transformation, and $s\in\{1,2,\ldots,r\}$.  Assume that
$\A$ has RIP with $\delta_{ts}<1$ for some positive constant $t$.
Then we have
\begin{eqnarray}\label{2f36} {\Gamma}_1(\A)\leq \frac{\sqrt{2}\delta_{ts}}{(1-\delta_{ts})\sqrt{t s-1}}.\end{eqnarray}
Furthermore, if $s<\frac{(1-\delta_{ts})\sqrt{ts-1}}{2\sqrt{2}\delta_{ts}}$, then ${\Gamma}_s(\A)\leq s{\Gamma}_1(\A)<1/2.$
\end{Theorem}
\textbf{Proof.} From Theorem \ref{ft8}, in order to establish the desired theorem, we only need to prove (\ref{2f36}). By Theorem \ref{ft4} and (\ref{2f29}), it is enough to show that for every $X\in\R^{m\times n}$ with $\A X=0$, we have\begin{eqnarray}\label{2f37} \|X\|=\|X\|_{1,\ast}\leq \hat{\gamma}\|X\|_\ast,~~~\hat{\gamma}:=\hat{\gamma}_1(\A)\leq \frac{\sqrt{2}\delta_{ts}}{(1-\delta_{ts})\sqrt{t s-1}}.\end{eqnarray}
As in the proof of Theorem \ref{ft9}, let SVD of $X$ be specified by
$$X=U\Diag(x)V^T,$$ where $U\in
\R^{m\times r}$ and $V\in \R^{n\times r}$, and
$\sigma(X):=x=(x_1,\ldots,x_r)^T$
is the vector of the singular values of $X$ with
$x_1\geq\cdots\geq x_r\geq 0$. Set $l=\lfloor ts/2 \rfloor$. We
decompose $x$ into a sum of vectors
$x_{T_i}, i\in\{0,1,2,\ldots\}$, where $T_0$ corresponds to the locations of the
largest entries of $X$, $T_1$ to the locations
of the next $l-1$ largest entries, and $T_j (j\geq 2)$ to the locations
of the next $l$ largest entries, and so on, with evident
modification for the last vector. We define
$X_{T_i}:=U\Diag(x_{T_i})V^T.$ Then, $X_{T_0}$ is the part of $X$ corresponding to the
largest singular values, $X_{T_1}$ is the part corresponding to the
next $l-1$ largest singular values, and $X_{T_j} (j\geq 2)$ is the part corresponding to the
next $l$ largest singular values, and so on. From the above partition, we easily
obtain that for $j\geq 2$, $$\|X_{T_j}\|_F\leq
l^{1/2}\|X_{T_j}\|\leq l^{-1/2} \|X_{T_{j-1}}\|_\ast.$$
Then it follows that
\begin{eqnarray} \sum_{j\geq 2}\|X_{T_j}\|_F\leq l^{-1/2}\sum_{j\geq 2} \|X_{T_{j-1}}\|_\ast\leq l^{-1/2}(\|X\|_\ast-\|X_{T_0}\|_\ast).\nonumber\end{eqnarray}
This yields
\begin{eqnarray}\|
X-X_{T_0}-X_{T_1}\|_F=\|\sum_{j\geq 2}X_{T_j}\|_F\leq\sum_{j\geq 2}\|X_{T_j}\|_F\leq l^{-1/2}(\|X\|_\ast-\|X_{T_0}\|_\ast)\leq l^{-1/2}\|X\|_\ast.\nonumber\end{eqnarray}
Together with $\A X=0$ and Lemma 3.3 \cite{CP09},
we obtain
\begin{eqnarray}0&=&\langle \A
(X_{T_0}+X_{T_1}), \A
X\rangle \nonumber\\
&= &\langle\A (X_{T_0}+X_{T_1}),\A (X_{T_0}+X_{T_1})\rangle+\langle\A (X_{T_0}+X_{T_1}),\A (X-X_{T_0}-X_{T_1})\rangle\nonumber\\
&\geq &(1-\delta_{l})\|X_{T_0}+X_{T_1}\|_F^2-l^{-1/2}\delta_{2l}\|X\|_\ast.\nonumber\end{eqnarray}
This implies
\begin{eqnarray}(1-\delta_{l})\|X_{T_0}+X_{T_1}\|_F&\leq& l^{-\frac{1}{2}}\delta_{ts}\|X\|_\ast.\nonumber\end{eqnarray}
Note the facts that $\|X\|_{1,\ast}=\|X_{T_0}\|_\ast \leq \|X_{T_0}\|_F\leq \|X_{T_0}+X_{T_1}\|_F$ and $\delta_l\leq\delta_{2l}\leq\delta_{ts}$ because of $l\leq ts/2$. We then have
\begin{eqnarray}(1-\delta_{l})\|X_{T_0}+X_{T_1}\|_F\leq(1-\delta_{ts})\|X_{T_0}+X_{T_1}\|_F\leq
\sqrt{\frac{2}{\delta_{ts}}}\delta_{ts}\|X\|_\ast\leq
\sqrt{\frac{2}{\delta_{ts-1}}}\delta_{ts}\|X\|_\ast.\nonumber\end{eqnarray}
This proves (\ref{2f37}) and hence the desired conclusion holds. {\qed}

\subsection{A bound for RIP}
From Theorems \ref{ft2} and \ref{ft9}, we actually provide a sufficient condition for
$s$-goodness in terms of RI constant $\delta_{2s}$: $\A$ is $s$-good if it
has the RIP with $\delta_{2s}<\sqrt{2}-1$. This establishes
a bound on the RI constant of $\A$.

\begin{Theorem} \label{fp3}  Let $b=\A W$ for some given $s$-rank matrix $W$. If
$\delta_{2s}<\sqrt{2}-1$, then $W=X^\ast$ where $X^\ast$ is the unique
optimal solution to NNM.
\end{Theorem}

Recht et al.
\cite{RFP08} showed that if $\delta_{5s}<1/10$, then $X^\ast=W$ where $X^\ast$ is the unique
optimal solution to NNM. Lee and Bresler \cite{LB09} gave
$\delta_{3s}<1/(1+4/\sqrt{3})$ by employing an analogue of the
approach for SSR \cite{Can08}; Cand\`{e}s and Plan \cite{CP09} gave
$\delta_{4s}<\sqrt{2}-1$ based on the work \cite{Can08,CT05}; Mohan
and Fazel \cite{MF10} gave $\delta_{2s}<0.307$, $\delta_{3s}<2\sqrt{5}-4$,
 and $\delta_{4s}<(8-\sqrt{40})/3$ by
combining a $s,s'$-restricted orthogonality constant property which
extended the recent work in CS \cite{CWX09,CWX10,CXZ09}.
Meka, Jain and Dhillon \cite{MJD09} gave
$\delta_{2s}<1/3$ via singular value projection (SVP), though
the  efficient SVP algorithm requires a priori knowledge of the rank
of $W$. Oymak, Mohan, Fazel and Hassibi \cite{OMFH2011}
proposed a general technique for translating results from SSR
to LMR, where they give the current best bound on
the restricted isometry constant $\delta_{2s}<0.472$.
Our results were independently obtained.

\section{Conclusion} In this paper, we studied the $s$-goodness characterization
of the linear transformation in LMR.
By employing the properties of $G$-numbers ${\gamma}_s$ and $\hat{\gamma}_s$,
we established necessary and sufficient conditions
for a linear transformation to be $s$-good, and provided sufficient conditions for exact and stable LMR via NNM under mild assumptions. Furthermore, we obtained computable upper bounds of ${G}$-number $\hat{\gamma}_s$, which lead
to verifiable sufficient
conditions for exact LMR.

\vskip4.5mm \noindent{\textbf{Acknowledgments} {The work was supported in part by the National
Natural Science Foundation of China (10831006)
and the National Basic Research Program of China (2010CB732501),
and a Discovery Grant from NSERC.}

%%%%%%%%%%%%%%%%%%%%%%%%%%%%%%%%%%%%%%%%%%%%%%%%%%%%%%%%%%%%
%%%%%%%%%%%%%%%%%%%%%%%%%%%%%%%%%%%%%%%%%%%%%%%%%%%%%%%%%%%%%%%%%%%%
 
\end{document}